%Paper: astro-ph/9308007
%From: tsvi@cfata4.harvard.edu (Tsvi Piran)
%Date: Thu, 5 Aug 93 12:01:19 -0400

\magnification 1200
%\baselineskip = 2\baselineskip
\def\ref{\par \smallskip \noindent \hangindent .5in \hangafter 1}
\def\etal{{\it et. al. }}
\centerline{\bf DO GAMMA-RAY BURST SOURCES REPEAT?}
\bigskip
\centerline{Ramesh Narayan and Tsvi Piran\footnote*{ Permanent Address:
Racah Institute for
Physics, The Hebrew University, Jerusalem 91904, Israel}}
\centerline{Harvard-Smithsonian Center for Astrophysics,}
\centerline{60 Garden Street, Cambridge, MA 02138, U.S.A.}
\bigskip

\centerline{\bf Summary}
\bigskip

Following the discovery by Quashnock and Lamb (1993) of an apparent
excess of $\gamma$-ray burst pairs with small angular separations, we
reanalyze the angular distribution of the bursts in the BATSE
catalogue.  We find that in addition to an excess of close pairs,
there is also a comparable excess of antipodal bursts, i.e pairs of
bursts separated by about 180 degrees in the sky.  Both excesses have
only modest statistical significance.  We reject the hypothesis put
forward by Quashnock and Lamb that burst sources are repeaters, since
it is obvious that this hypothesis does not predict an excess of
antipodal coincidences.  Lacking any physical model of bursts that can
explain the antipodal pairs, we suggest that the two excesses seen in
the data are either due to an unusual statistical fluctuation or
caused by some unknown selection effect.
\bigskip
{keywords: Gamma-rays: bursts}
%\vfill\eject

\bigskip\bigskip
\noindent
{\it 1. Introduction}

In a recent paper Quashnock and Lamb (1993, denoted QL) analyze the
distribution of angular separations of bursts in the publicly
available BATSE catalogue of gamma-ray bursts.  They find a
significant excess of close pairs of bursts with angular separations
smaller than $\sim 4^o$, compared to the number of such pairs expected
for a random distribution of positions on the sky.  On this basis they
suggest that $\gamma$-ray bursters repeat.  In their hypothesis, the
close pairs actually arise from the same source, but they are assigned
slightly different positions because of measurement errors, which are
typically about $4^o$ or larger.  If QL are correct, then many
extragalactic models may be ruled out; in particular, the neutron star
merger model (Eichler et al, 1989; Narayan, Paczynski and Piran 1992)
would become rather unlikely.  Furthermore, as QL argue, their result
implies a close relationship between classical $\gamma$-ray bursts and
the three soft $\gamma$-ray repeaters.  Since the latter are known to
be located close to or in the Galaxy, this would give further support
to the suggestion that the classical burst sources too are located in
the Galaxy.  In view of the importance of these conclusions we
reanalyze in this letter the angular distribution of gamma ray bursts.

We repeat here the nearest neighbour analysis employed by QL and add to
it another analysis based on the more standard angular autocorrelation
function.  We find that, while there does appear to be an excess of
close pairs of bursts with angular separations less than $4^o$ as
claimed by QL, there is also an excess of antipodal pairs of bursts
with angular separations larger than $176^o$.  QL explain the excess
of nearby bursts as signals from repeating sources, but we cannot
think of any physical model to explain the excess of antipodal bursts.
We thus conclude that both effects, if real, are most likely due to
some unknown selection effect.

In sec. 2 of this letter we analyze the data.  We examine the
correlation function of the full sample of 260 bursts in the BATSE
catalogue in sec. 2.a. and we present the nearest and farthest
neighbour analysis for this sample in sec. 2.b.  QL do not use the full
sample, but analyze a subsample defined by the avaliability of counts
in both the 64 ms and the 1024 ms channels on BATSE.  Another subset
could be defined by including only those bursts for which the formal
positional accuracy is better than $4^o$.  We discuss results from
these subsamples in sec. 2.c.  We conclude in sec. 3 with a discussion
of the implications of the results.

\bigskip
\noindent
{\it 2. Analysis}
\bigskip
\noindent
{\it 2.a The Two Point Angular Correlation Function}
\bigskip

A data set composed of randomly positioned sources, some of which
repeat, has a simple angular correlation function: a delta function
peak at the origin due to the repeaters, and a slightly negative
constant elsewhere.  Errors in position measurements will spread out
the delta function to a broadened peak at the origin with a width and
shape corresponding to the probability distribution of the position
errors, but nowhere else do we expect any other significant peak or
dip.  This simple structure of the two-point correlation function
suggests that it ought to be a clean statistic to test the repeating
source hypothesis.

The correlation function of all the 260 bursts in the BATSE catalogue
is shown in Fig. 1(a).  As expected from the QL analysis, there is a
peak in the bin corresponding to $\theta < 4^o$.  To assess the
statistical significance of this peak we have carried out Monte Carlo
simulations with 10000 random samples.  From this we estimate that the
peak has an amplitude of 1.75 standard deviations $(1.75 ~\sigma$),
and that the probability of obtaining a peak as strong as or stronger
than the observed one is 0.0957 (9.57\%).  These results are given in
Table 1 along with the results of various other calculations discussed
later in the paper.  Note that the probability distribution of the
values of the correlation function is not Gaussian.  Therefore, here
and elsewhere in the text, all significance levels that we quote are
obtained from Monte Carlo simulations.  Note also that a small
probability in Table 1 means that it is unlikely for the particular
event to have happened by chance and therefore implies high
significance.

While the existence of a peak at $\theta <4^o$ agrees with the QL
result, there is another unexpected peak at the antipode,
corresponding to the bin with $\theta > 176^o$.  The second peak is
slightly wider than the peak at the origin and has an amplitude of
$1.86~\sigma$, which is marginally more significant (probability $=
0.0724$) than the direct peak.  In fact, while the statistical
significance of each of the single peaks may be considered marginal,
the statistical significance of having {\it both} peaks is much
higher.  Additionally, there are weak negative excursions in the
correlation function at around $\theta \sim10-20^o$ and $\theta \sim
160-170^o$ which appear to be moderately significant, but we do not
discuss these in detail.

\vfill\eject
%\bigskip\bigskip
\noindent
{\it 2.b Nearest Neighbour Analysis.}
\bigskip

Quashnock and Lamb did not employ the correlation function but instead
used a ``nearest neighbour statistic.''  According to this statistic
they measure for each burst in the catalogue the angular distance to
its nearest neighbour and compute the cumulative distribution of this
quantity for the observed sample.  They then compare this cumulative
with the distribution expected for a random sample.  In Fig. 1(b) we
present this comparison for the full sample of 260 bursts.  We show
the number of bursts with nearest neighbours closer than an angle
$\theta$ as a function of $1- \cos\theta$.  As expected from QL's
analysis and from the correlation function approach discussed in sec.
2.a., there is evidence for an excess of nearby bursts.  Motivated by
our discovery of the antipodal peak in the correlation function, we
also calculate for each burst the distance to its {\it farthest}
neighbour, that is the burst that is nearest to the antipodal point.
We plot this cumulative distribution in Fig 1(b), where now the x-axis is
$1+\cos\theta$ and the vertical axis represents the number of bursts
with farthest neighbour more distant than angle $\theta$.  As expected
from the correlation function, we find that there is again an excess
of antipodal bursts compared to the distribution expected for random
bursts.  Visually at least, we would say that the evidence for an
excess is about equally strong for nearest and farthest neighbours.

What is the significance of these deviations?  Since Fig. 1(b) is a
cumulative distribution, one might consider using the
Kolmogorov-Smirnov (KS) test to estimate the significance.  In this
method one measures the KS distance, which is the maximum distance
between the observed cumulative curve and the theoretically expected
curve.  One then calculates the probability of obtaining a distance at
least as large as the observed one by the standard KS method (e.g.
Press et al, 1992).  However, the KS method assumes that all the data
points are independent, but this is not the case in the present
distribution because the set of nearest neighbour distances can have
strong correlations.  To see this, consider the case when two bursts
happen to lie very close to each other on the sky.  Both bursts will
yield the same small nearest neighbour distance, and therefore this
distance will be counted twice.  This demonstrates that a direct
application of the KS method is invalid.

To measure correctly the statistical significance of the deviations we
generated 10000 random data sets and calculated the distribution of
the KS distance.  We find that the probability of obtaining KS
distances greater than those observed in the 260 burst BATSE sample
are: 0.012 for the nearest neighbours and 0.11 for the farthest
neighbours.  (Note that the standard KS test gives probabilities of
0.0016 and 0.019, showing that the KS test, by neglecting the effect of
correlations, leads to unduly optimistic estimates of the
significance.)  The results are marginally significant at best.  If we
estimate the probability for both nearest and farthest neighbours to
have such large deviations simultaneously, we find that it is 0.0019,
which is somewhat more significant.  Therefore, once again we reach
the same conclusion as we did from the correlation analysis in sec.
2.a, viz. while we find an excess of nearest and farthest neighbours
in the data, these excesses are only marginally significant.  However,
the probability of obtaining both excesses simultaneously is very
small and therefore the evidence for such a signal is more
significant.

\bigskip
\noindent
{\it 2.c Subsamples of the BATSE Catalogue}
\bigskip

The basic variables in our analysis are the positions of the bursts.
As stated in the instructions with the BATSE catalogue, burst
positions have variable errors depending on the strengths of the
bursts, the position and orientation of the Compton GRO satellite, and
other parameters.  The BATSE catalogue gives an estimate of the formal
positional error $\Delta \theta _P$ for each burst due to Poisson
fluctuations in the observed $\gamma$-ray counts, to which an
additional systematic error of $4^o$ should be added in quadrature to
yield the total positional error, $\Delta \theta _{tot} =
[\Delta\theta _P^2+(4^o)^2]^{1/2}$.  In some cases the estimated
$\Delta \theta _P$ is quite large (up to $20^o$) and it is reasonable
to exclude such bursts from the samples.  Since we are looking for
correlations on angular separations of $4^o$ or less, it is clearly
meaningless to argue that a burst with a positional error of say,
$10^o$, is within $4^o$ from another burst.  We have therefore
repeated our analysis with a sub sample of bursts for which the quoted
formal $\Delta\theta _P$ values are $4^o$ or less.  This reduces the
sample from 260 bursts to 131 bursts.  When we analyze this sample,
the correlation analysis gives a modest peak of $0.84 \sigma$ for
$\theta <4^o$, and an impressive peak of $2.69 \sigma$ for antipodal
neighbours between $176^o$ and $180^o$.  The corresponding
probabilities for chance occurrence are: $ 0.42$ for the direct peak
and $ 0.010$ for the antipodal peak (see Table 1).  Similarly, the QL
statistic based on the KS distance gives excesses of $0.83\sigma$ and
$1.61\sigma$ for nearest and farthest neighbours, corresponding to
chance probabilities of $0.47$ and $0.027$. The fact that the
statistical significance of the effect does not increase, and in fact
{\it decreases}, when we throw away the bursts with large positional
errors supports our suspicion that the observed anomaly is not due to
a real physical effect.

For completeness, we have also repeated the analysis with the
particular sub-sample used by QL.  They divide the data into subgroups
according to the $\gamma$-ray counts measured in the 64 ms and 1024 ms
channels.  Their total sample, identified as Type I+II in their paper,
consists of 201 bursts; we refer to this as the QL sample.  Among all
the subsamples that QL analyzed, they found the strongest signal in
this particular combined sample.  Using 10000 Monte Carlo simulations
with synthetic data, we find that there is a probability of 0.0015 of
obtaining by chance a nearest neighbour deviation comparable to the
signal observed in the QL sample.  We also find a modest excess of
farthest neighbours in this sample, with a Monte Carlo probability of
0.36.

Our final subset is a truncated QL sample of 108 bursts, where we
take the QL sample and eliminate all bursts with $\Delta
\theta _P >4^o$.  The hypothesis of repeating bursts put forward by QL
suggests that by eliminating bursts with highly uncertain positions
the statistical significance of the effect should increase.  We find
this not to be the case.  The angular correlation function for this
data set is shown in Fig. 2(a). We see both the forward and antipodal
peaks, but the forward peak represents only a $.74\sigma$ deviation
while the antipodal peak is at $2.60\sigma$ (see Table 1 for the
corresponding probabilities).  Similarly, we show in Fig. 2(b) the
nearest/farthest statistic.  Here we find that the KS distance for the
distribution of nearest bursts is only $0.68\sigma$ (random chance
probability of 0.66), while for the farthest bursts the KS distance is
$1.24\sigma$ corresponding to a random chance probability of $0.10$.

\bigskip
\noindent
{\it 3. Summary and Discussion}
\bigskip
Our primary conclusions are the following:

\noindent
(i) The BATSE data does contain some evidence for an excess of pairs
of bursts with angular separations smaller than $4^o$.  However, there
is equally good evidence for an excess of nearly antipodal pairs, with
separations between $176^o$ and $180^o$.  An antipodal peak is, of
course, not expected from a random population of repeaters.

\noindent
(ii) The statistical significance of either excess is marginal.  We
find that the only statistic that provides a reasonably strong signal
consistently is when we combine the excesses in the forward and
antipodal directions and compute the probability of obtaining both
simultaneously.  There are obvious dangers in carefully selecting two
hypotheses in this fashion and combining them, therefore we are
ambivalent about accepting the combined excess as a real property of
the BATSE sample of bursts.

\noindent
(iii) When we eliminate bursts with Poisson positional errors $\Delta
\theta_P >4^o$ (which corresponds to total error $\Delta \theta
_{tot}>5.7^o$), far from becoming stronger, the evidence for the
excesses actually becomes weaker.  In fact the decrease in the signal
is quite drastic in the case of nearest neighbour pairs, while it is
more modest for the antipodal pairs.  In other words, the evidence for
an excess of close pairs is very volatile, depending on the particular
sample chosen, while the excess of antipodal pairs shows a little more
stability at least within the tests we have done (Table 1).  Note that
bursts with large $\Delta\theta_P$ are generally weaker and one could
argue that by eliminating them we eliminate the repeating weak bursts
(according to the QL model).  However, the large positional error of
these weak bursts means that these bursts should not have shown any
evidence for the repeater model in the first place.

We have been unable to come up with any physical model that can
explain an excess of bursts within $4^o$ of the antipode.  Such an
excess may occur if the bursts are located along a narrow line in
space or in a very thin disk.  However, such distributions are clearly
ruled out by the observed overall isotropy of the positions of the
bursts (Meegan et al, 1992).

We are forced to conclude that the excess of close pairs of bursts
discovered by QL, and the antipodal excess that we discuss here, are
either due to a statistical fluctuation or caused by some selection
effect{\footnote{\P}{The possibility of a selection effect is
supported by the fact that the two bursts identified by trigger
numbers 803 and 974 in the BATSE catalogue have identical positions to
all digits given.  However, one of the bursts has a positional
accuracy of only $16^o$, making the agreement of the two positions
highly unlikely.}}.  We do not have any specific idea as to the nature
of the selection effect.

We thank Dalia Goldwirth for useful discussions, and Don Lamb and Jean
Quashnock for help in understanding their burst samples and for
comments. This work was supported by NASA grant NAG 5-1904.
%\vfill\eject
\bigskip\bigskip
\noindent
{\it References}
\ref
Eichler, D., Livio, M., Piran, T. \& Schramm, D.N. 1989, Nature, {\bf 340},
126.
\ref
Meegan, C.A., \etal 1992, Nature, {\bf 355} 143.
\ref
Narayan, R., Paczy\'nski, B. \& Piran, T. 1992, ApJL, {\bf 395}, L83.
\ref
Press, W.H., Teukolsky, S.A., Vetterling, W.T. \& Flannery, B.P.
1992, {\it Numerical Recipies}, Cambridge University Press,
Cambridge.
\ref
Quashnock, J. Q \& Lamb, D. 1993, MNRAS, submitted.
%\vfill\eject
\bigskip\bigskip
\centerline {\bf Table I- Correlations functions and significance levels}
$$\vbox{\tabskip=0pt \offinterlineskip
\halign to\hsize{\strut#
\tabskip=1em plus2em&\hfil#& #& \hfil#& #
&\hfil#& #&
&\hfil#& #&\hfil#& # &\hfil#& #
\tabskip=0pt\cr \noalign{\ }
&&  \ Sample &&    Number
&& $\ \ \   c(\le 4^o) \ \ \ $ &&   ~Prob && $ \ \ \ c(\ge 176^o)\ \ \ $
&& ~Prob &\cr \noalign{\ }
&&      &&    && &&  && &&   &\cr \noalign{\ }
&&   Full Sample   && \ 260   && $0.268=1.75\sigma$
&&  0.096    && $0.292=1.86\sigma$  &&  0.072
&\cr \noalign{\ }
&&   QL  Sample   && \ 201    && $0.389=1.95\sigma$
&&  0.052    && $0.348=1.74\sigma$  && $ 0.082 $
&\cr \noalign{\ }
&&   Full, $ \Delta \theta_P\leq 4^o$    && \ 131   && $ 0.254=0.84\sigma$
&&    0.423     && $0.832=2.69\sigma$  &&  0.010
&\cr \noalign{\ }
&&    QL, $ \Delta \theta_P\leq 4^o$  && \ 108   && $ 0.279=0.74\sigma$
&&  0.575   && $0.989=2.60\sigma$  &&  0.016
&\cr \noalign{\ }
\noalign{\smallskip} \hfil\cr}}$$
%\vfill\eject
\bigskip\bigskip
\noindent
{\it Figure Captions}
\ref
{\it Fig. 1: (a)} The angular autocorrelation function of the full
sample of 260 bursts in the BATSE catalogue, represented in $4^o$
bins.  The error bars correspond to one standard deviation as
estimated through Monte Carlo simulations.  Note that the antipodal
peak near $\theta \sim 180^o$ is somewhat higher and wider than the
direct peak near $\theta\sim 0^o$.  There are marginally significant
negative dips adjascent to both peaks and it is plausible that the
excess counts in the peaks are supplied by the lack of bursts in the
dips.  {\it (b)} The cumulative number of nearest neighbours closer
than $\theta$ as a function of $1-\cos \theta$ and farthest neighbours
more distant than $\theta$ as a function of $1+\cos \theta$, both
shown as thick lines.  The theoretically expected curve for a random
sample is shown by the smooth thin line.  Note that the excess extends
to larger angles in the case of farthest neighbours, whereas the KS
statistic (i.e. the largest distance between the observed and
theoretical curves) is slightly higher in the case of nearest
neighbours.

\ref
{\it Fig. 2: (a)} The angular autocorrelation function of the 108
bursts out of the QL sample which have Poisson position errors
$\Delta\theta _P \leq 4^o$.  Note that the antipodal peak is much
stronger than the direct peak, the latter being practically
insignificant.  {\it (b)} The cumulative number of nearest neighbours
as a function of $1-\cos \theta$ and farthest neighbours as a function
of $1+\cos \theta$, shown as thick lines.  The theoretically expected
curve is shown by the thin line.  The nearest neighbour curve is very
close to the expected curve whereas the farthest neighbour curve shows
an apparently significant deviation.

\end